\documentclass[fleqn,usenatbib]{mnras}
\usepackage{newtxtext,newtxmath}

\usepackage[T1]{fontenc}
\usepackage{ae,aecompl}

\usepackage{lastpage, graphicx, amssymb, aas_macros, bm, nicefrac}

\usepackage{bigints}
\usepackage{tabularx}
\usepackage{colortbl}
\usepackage{dcolumn}
\usepackage{xcolor}
\usepackage{ulem}
\usepackage{url}
\hypersetup{breaklinks, colorlinks=true, urlcolor=blue, citecolor=blue, linkcolor=blue}
\newcommand{\mstarmax}{M_{\star, \rm{max}}}
\newcommand{\fbary}{f_{\rm{b}}}

\definecolor{Gray}{gray}{0.9}
\definecolor{texas}{HTML}{BF5700}
\definecolor{update}{HTML}{800000}

\newcommand{\mstar}{M_{\star}}
\newcommand{\msun}{M_{\odot}}

\newcommand{\mpc}{{\rm Mpc}}

\newcommand{\kms}{{\rm km \, s}^{-1}}

\newcommand{\lcdm}{$\Lambda$CDM}

\newcommand{\jwst}{\textit{JWST}}

\newcommand{\mhalo}{M_{\rm halo}}

\renewcommand{\ga}{\gtrsim}

\defcitealias{labbe2022}{L22}

\title[Ruling out \lcdm\ with high-redshift galaxies (?)]
{
Stress testing \lcdm\ with high-redshift galaxy candidates}
\author[M. Boylan-Kolchin]
{Michael Boylan-Kolchin\\
\noindent $\!\!$Department of Astronomy, The University of Texas at Austin,
2515 Speedway, Stop C1400, Austin, TX 78712-1205, USA; \url{mbk@astro.as.utexas.edu}}

\pubyear{2023}

\begin{document}
\label{firstpage}
\pagerange{\pageref{firstpage}--\pageref{LastPage}}
\maketitle

\begin{abstract}
Early data from \jwst\ have revealed a bevy of high-redshift galaxy candidates with unexpectedly high stellar masses. An immediate concern is the consistency of these candidates with galaxy formation in the standard cosmological model. In the \lcdm\ paradigm, the stellar mass ($\mstar$) of a galaxy is limited by the available baryonic reservoir of its host dark matter halo. The mass function of dark matter halos therefore imposes an absolute upper limit on the number density $n(>\mstar,z)$ and stellar mass density $\rho_{\star}(>\mstar,z)$ of galaxies more massive than $\mstar$ at any epoch $z$. Here I show that the most massive galaxy candidates in \jwst\ observations at $z\sim 7-10$ lie at the very edge of these limits, indicating an important unresolved issue with the properties of galaxies derived from the observations, how galaxies form at early times in \lcdm, or within this standard cosmology itself. 
\end{abstract}


\section{Introduction}
\lcdm-like cosmological models share a similar basic assumption: baryons and dark matter are well-mixed at very early times, and as baryons collapse into dark matter halos, the maximum amount of baryonic material within a halo will be equal to $M_{\rm b}=\fbary\,\mhalo$, where $\fbary \equiv \Omega_{\rm b}/\Omega_{\rm m}$ is the cosmic baryon fraction. This, in turn bounds the total stellar content of a dark matter halo: $\mstar(\mhalo) \leq M_{\rm b}(\mhalo)$. I show how this simple relation can be used as a stringent test of either cosmological models with minimal assumptions about galaxy formation or the reliability of photometric selection and physical characterization of high-redshift galaxy candidates. My analysis is in many ways similar to \citet{behroozi2018}, who connected cumulative number densities of dark matter halos to high-redshift galaxy stellar mass functions (see also \citealt{steinhardt2016}), though I also consider the maximal cumulative stellar mass density allowed in \lcdm. The question of the consistency of stellar mass functions and the underlying cosmological dark matter halo mass functions has become considerably more urgent with the release of the first data from \jwst, and with it, a swarm of high-redshift galaxy candidates \citep{adams2023, atek2023, castellano2022, donnan2023, finkelstein2022, labbe2022, morishita2022, naidu2022, yan2023}. 

\begin{figure*}
 \centering
 \includegraphics[width=0.49\textwidth]{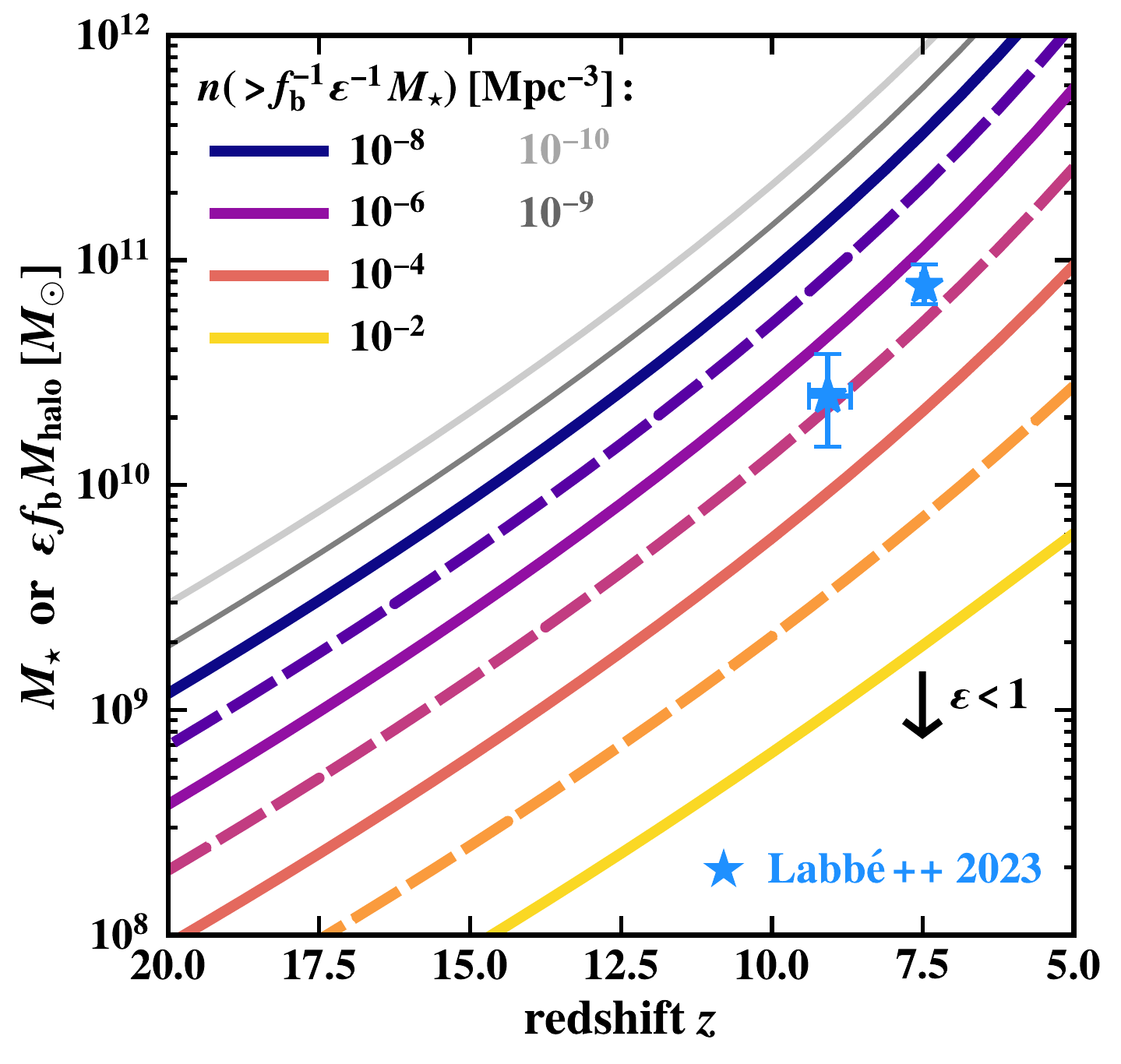}
  \includegraphics[width=0.49\textwidth]{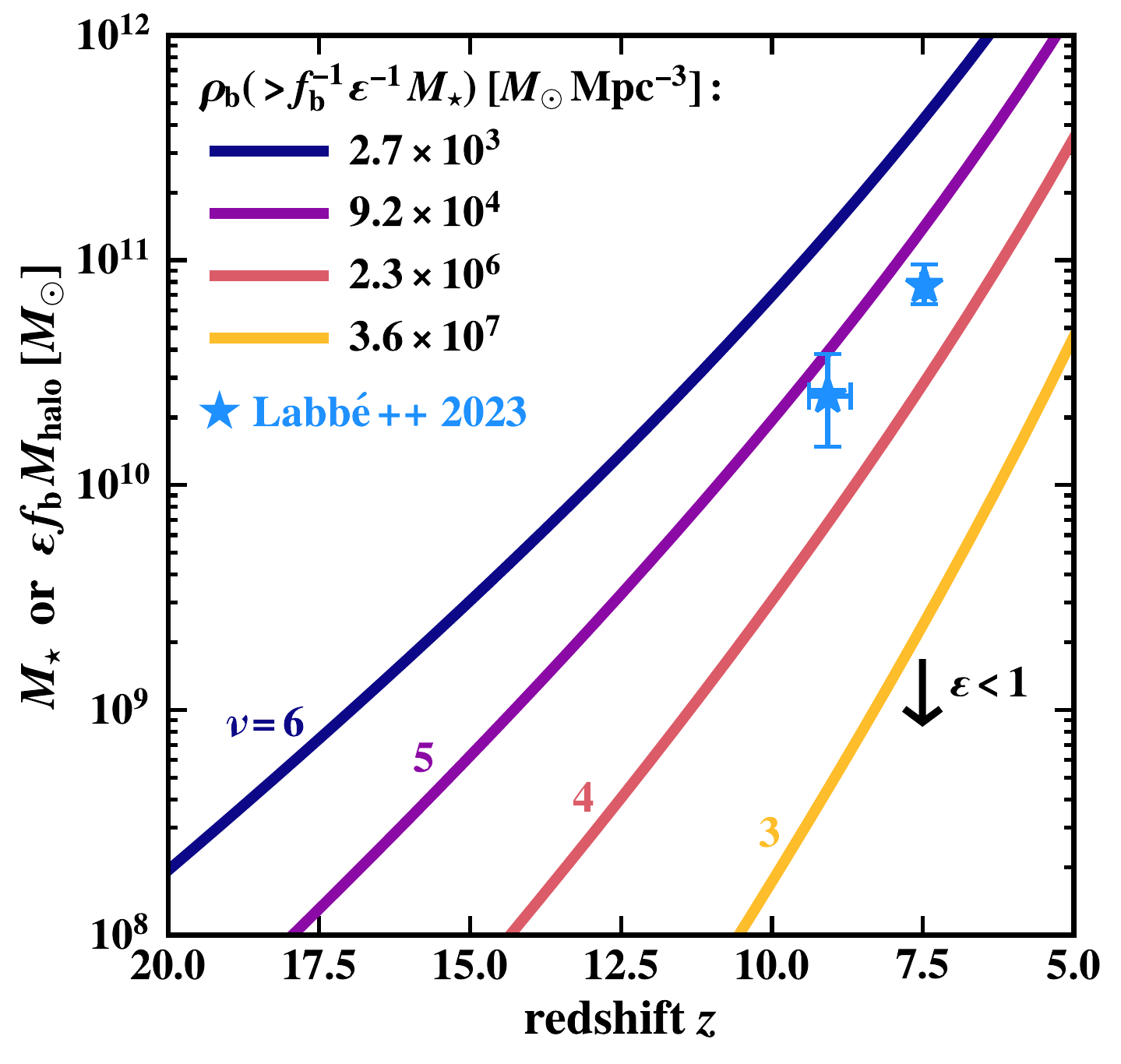}
 \caption{\textit{Limits on the abundance of galaxies as a function of redshift.} Curves show the relationship between $\mstar$ and $z$ at fixed cumulative halo abundance (\textit{left}) and fixed $\rho_{\rm b}(>\mhalo)$, or equivalently fixed peak height $\nu$ (\textit{right}). The most extreme \citetalias{labbe2022} galaxy candidates are shown as blue stars, with uncertainties indicating 68\% intervals (symmetric about the median) of the posterior probability distribution.
  The existence of a galaxy with $\mstar$ at redshift $z$ requires that such galaxies have a cumulative comoving number density that is at most the number density shown in the left panel, as those galaxies must reside in host halo of mass $\mhalo=\mstar/(\fbary\,\epsilon)$. The cumulative comoving number density corresponding to an observed $\mstar$ will likely be (much) smaller than is indicated here, as the curves are placed on the plot by assuming the physically maximal $\epsilon=1$. For smaller values of $\epsilon$, the curves in each panel move down relative to the points by a factor of $\epsilon$ (as indicated by black downward-facing arrows). The right panel demonstrates that even for the most conservative assumption of $\epsilon=1$, the data points correspond to very rare peaks in the density field, implying a limited baryonic reservoir that is in tension with the measured stellar masses of the galaxies.
}
 \label{fig:fig1}
\end{figure*}

\vspace{-0.2cm}
\section{Assumptions}
I adopt the base \lcdm\ model of \citet{planck2020}, which assumes no spatial curvature and initial conditions that are Gaussian and adiabatic, as the standard cosmological model. I use best fit values for cosmological parameters based on the {\tt Plik} TT,TE,EE+lowE+lensing likelihood applied to the full-mission data. The relevant parameters and values for this work are the present-day Hubble constant, $H_0=67.32\,\kms\,\mpc^{-1}$; the $z=0$ density parameter for matter, $\Omega_{\rm m}=0.3158$ (which includes baryons, dark matter, and non-relativistic neutrinos); the slope of the primordial power spectrum of density fluctuations, $n_{\rm s}=0.96605$; the rms amplitude of the linear matter power spectrum at $z=0$ as measured in spheres of radius $8\,h^{-1}\,\mpc$, $\sigma_{8}=0.8120$; and the cosmic baryon fraction, $\fbary \equiv \Omega_{\rm b}/\Omega_{\rm m}=0.156$~\citep{planck2020}.

With these values, the linear matter power spectrum is specified at all times relevant for structure formation. The non-linear density field, home to the dark matter halos that host galaxies, must be computed numerically. However, a long line of research starting with \citet{press1974} has been devoted to connecting the abundance of dark matter halos as a function of redshift and mass to the underlying linear matter power spectrum. In what follows, I use the \citet{sheth1999} dark matter halo mass function $dn(M,z)/dM$ --- the number of dark matter halos of mass $M$ per unit mass per unit comoving volume at redshift $z$ --- to compute the comoving number density of halos above a given halo mass threshold, 
\begin{equation}
    \label{eq:n_gt_m}
    n(>\mhalo,z)=\int_{\mhalo}^{\infty} dM\,\frac{dn(M,z)}{dM}\,
\end{equation}
and the comoving mass density in halos more massive than $\mhalo$, 
\begin{equation}
    \label{eq:rho_gt_m}
    \rho_{\rm m}(>\mhalo,z)=\int_{\mhalo}^{\infty} dM\,M\,\frac{dn(M,z)}{dM}\,.
\end{equation}
These translate directly to upper limits on the statistics of galaxies through the straightforward assumption that the largest stellar content a halo can have given its cosmic allotment of baryons is $\mstarmax = \fbary\,\mhalo$. More generally, we may write $\mstar=\epsilon\,\fbary\,\mhalo$, with $\epsilon \leq 1$ being the efficiency of converting baryons into stars. 

The cumulative comoving number density of dark matter halos more massive than $\mhalo$ thus sets an upper limit on the comoving number density of galaxies more massive than $\mstar$,  
\begin{equation}
\label{eq:n_gt_mstar}
    n_{\rm gal}(> \mstar) \leq n_{\rm halo}(>\mstar/\fbary)\,.
\end{equation}
Similarly, the cumulative comoving density of collapsed mass  sets an upper limit on the density of collapsed baryons, $\rho_{\rm b}(> \mhalo) = \fbary\,\rho_{\rm m}(>\mhalo)$, which in turn strictly bounds the comoving mass density of stars contained in halos more massive than $\mhalo$, \begin{equation}
\label{eq:rhob_gt_mhalo}
    \rho_{\star}(> \mhalo) \leq \fbary\,\rho_{\rm m}(>\mhalo)\,,
\end{equation}
and the density of stars contained in galaxies above a given $\mstar$, 
\begin{equation}
\label{eq:rhob_gt_mstar}
    \rho_{\star}(> \mstar) \leq \fbary\,\rho_{\rm m}(>\mstar/\fbary)\,.
\end{equation}

\section{Results}
\begin{figure*}
 \centering
 \includegraphics[width=0.49\textwidth]{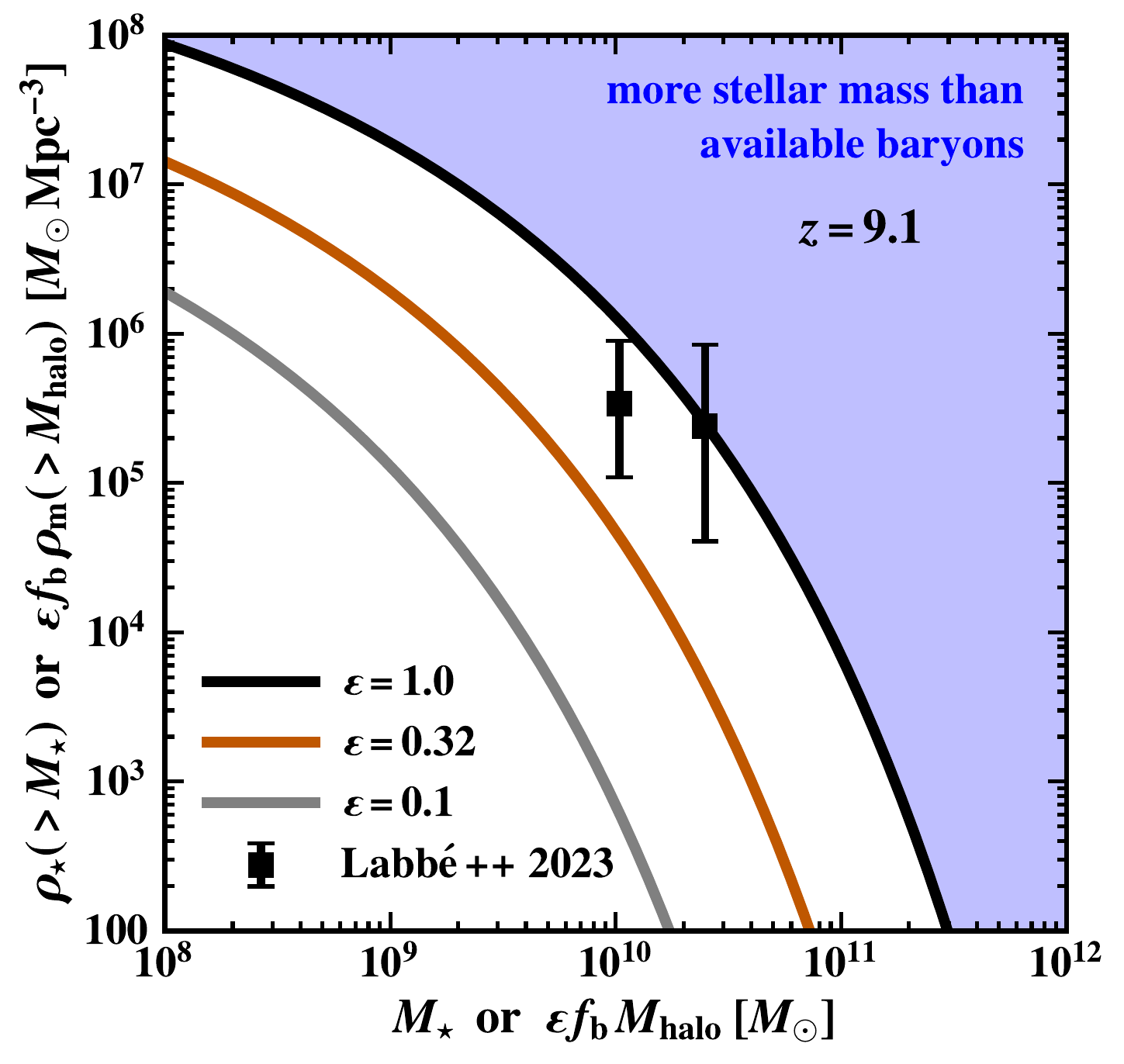}
  \includegraphics[width=0.49\textwidth]{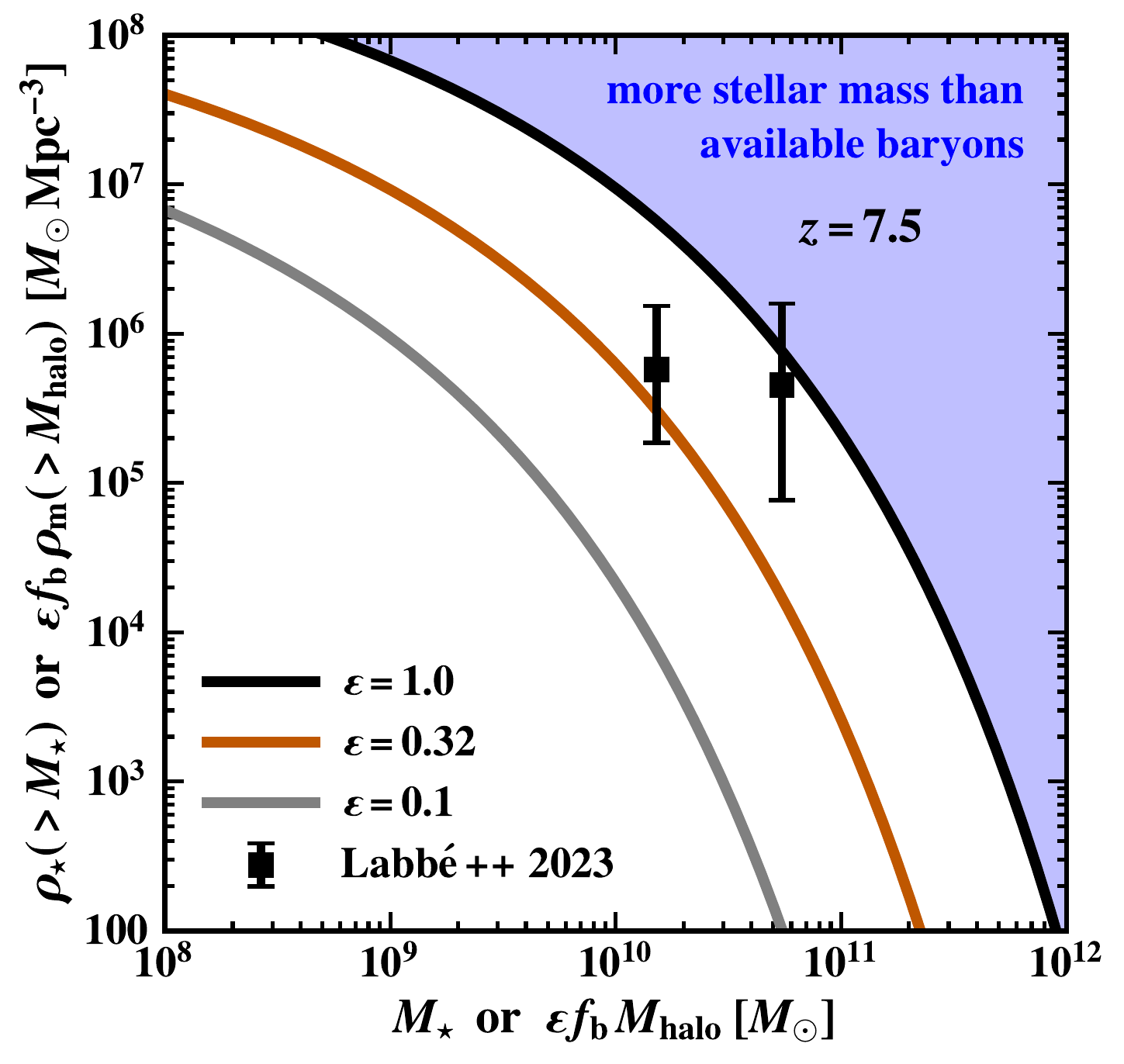}
 \caption{\textit{Stellar mass density limits.}
 The comoving stellar mass density contained within galaxies more massive than $\mstar$ at $z\approx 9.1$ (\textit{left}) and $z\approx 7.5$ (\textit{right}) for three values of the assumed conversion efficiency $\epsilon$ of a halo's cosmic allotment of baryons into stars. Only if \textit{all} available baryons in all halos with enough baryons to form the galaxies reported by \citetalias{labbe2022} have indeed been converted into stars by that point --- an unrealistic limit --- 
 is it possible produce the stellar mass density in the highest $\mstar$ bin at $z\approx 9$ measured by \citetalias{labbe2022} in a typical volume of a \lcdm\ Universe with the Planck 2020 cosmology. Results are similar at $z \approx 7.5$. For more realistic values of $\epsilon$, the required baryon reservoir is substantially larger than the theoretical maximum in this cosmology. 
 When considering shot noise and sample variance errors (which comprise the plotted uncertainties on the \citetalias{labbe2022} data points in each panel), the measurements are consistent with the base \lcdm\ model if $\epsilon > 0.57$, which would still imply incredibly efficient star formation in the high-redshift Universe.
 }
 \label{fig:rho_star}
\end{figure*}

The left panel of Figure~\ref{fig:fig1} shows the relationship between the maximal inferred stellar mass for a given $\mhalo$, $\mstar=\fbary\,\mhalo$ (i.e., assuming the maximal $\epsilon=1$), and redshift $z$ for fixed cumulative comoving halo number densities ranging from $10^{-10}\,\mpc^{-3}$ (light gray) to $10^{-2}\,\mpc^{-3}$ (yellow). The curves evolve rapidly with redshift, with the maximal stellar mass corresponding to a fixed cumulative comoving halo number density increasing by two orders of magnitude from $z=20$ to $z=8$. This rapid rise indicates that the mass reservoir available for the most massive galaxies increases quickly with redshift at fixed halo number density. The two most massive high-redshift galaxy candidates from the \citet[hereafter \citetalias{labbe2022}]{labbe2022} sample, at $z\approx 7.5$ ($\mstar \approx 10^{11}\,\msun$) and $z \approx 9.1$ ($\mstar \approx 10^{10.5}\,\msun$), are shown as blue stars. These objects are unexpectedly massive, with stellar content reflective of halos that have cumulative comoving number densities no higher than $\approx 10^{-5.2}\,\mpc^{-3}$ (if $\epsilon=1$); for $\epsilon=0.32$ (0.1), the implied number density is $\approx 10^{-7}$ ($10^{-9.3}$)~$\mpc^{-3}$. By comparison, the candidates were found in a survey of 38 arcmin$^2$, a volume of $V \approx 10^{5}\,\mpc^{3}$ at each of the redshift bins --- $7 < z < 8.5$ and $8.5 < z < 10$ --- considered by \citetalias{labbe2022}.

The right panel of Figure~\ref{fig:fig1} recasts the issue in terms of the scarcity of systems as measured by cumulative mass density. In extended Press-Schechter models, the peak height $\nu(\mhalo, z)=\delta_{\rm c}/\sigma(\mhalo,z)$ of an object --- where $\delta_{\rm c} \approx 1.7$ is the collapse threshold and $\sigma^2(\mhalo,z)$ is the variance of the linear density field at redshift $z$ smoothed on a scale containing an average mass of $\mhalo$ --- is a measure of the fraction of mass in the Universe contained in virialized objects more massive than $\mhalo$ at redshift $z$. Typical halos at $z$ have $\nu=1$, which corresponds to 24\% of the mass in the Universe residing in halos at least that massive; larger values of $\nu$ indicate increasingly massive and therefore rare peaks in the density field at that epoch. The comoving baryon density for each peak height in the figure is given in the legend; multiplying this number by the volume of a survey gives the total amount of baryons contained above the mass corresponding to that peak height and redshift. The \citetalias{labbe2022} galaxies have peak heights of at least $\nu=4.5$ (assuming $\epsilon=1)$, meaning that at most a fraction $6.2 \times 10^{-5}$ of the baryons in the Universe are contained in halos massive enough to host these galaxies. For reference, $\nu=4.5$ at $z=0$ corresponds to $\mhalo \approx 5 \times 10^{15}\,\msun$. Adopting more reasonable efficiencies of $\epsilon=0.32$ or $0.1$ results in rarer peaks with $\nu \approx 5.4$ or $6.4$.

Figure~\ref{fig:rho_star} shows the cumulative stellar mass density reported by \citetalias{labbe2022} at $z\approx 9$ (left) and $z \approx 7.5$. The data, which come from individual massive objects, lie at the extreme of \lcdm\ expectations even in the most optimistic scenario: at both redshifts, the measurements lie at the theoretical limit of $\rho_{\star}(>\mstar)=\fbary\,\rho_{\rm m}(>\mstar/\fbary)$, implying physically implausible values of $\epsilon(z \approx 9)=0.99$ and $\epsilon(z \approx 7.5)=0.84$.
When considering the $1\,\sigma$ error (which incorporates uncertainties in the stellar mass estimation, Poisson fluctuations, and sample variance), the data become marginally consistent with the available baryon reservoirs for an efficiency of $\epsilon(z \approx 9) \ge 0.57$, which is likely an unrealistically high value. Assuming a more plausible value of $\epsilon=0.1$ or $0.32$ yields a strong discrepancy with \lcdm\ expectations at both redshifts even when considering observational uncertainties.

\section{Discussion}
\label{sec:discussion}
The first glimpse of high-redshift galaxy formation with \jwst\ has revealed surprisingly massive galaxy candidates at early cosmic times. These systems provide a way to test a bedrock property of the \lcdm\ model (or, e.g., assumptions in derivations of stellar masses or the viability of high-redshift galaxy candidates): the stellar content of halos should not exceed the available baryonic material in those halos. This requirement does not rely on assumptions such as abundance matching but rather is simply a statement about the distribution of virialized mass in the Universe as a function of redshift and the baryonic reservoirs associated with those virialized halos: galaxies of mass $\mstar$ can only form if halos of mass $\mstar/(\epsilon\,\fbary)$ have formed. It is also more stringent than the requirement that the observed galaxy UV luminosity function not exceed the theoretical maximum coming from a nearly instantaneous (10 Myr) conversion of a halo's full baryonic reservoir into stars \citep{mason2022}, as it is an integral constraint as opposed to a differential one. The massive, high-redshift galaxy candidates cataloged in \citetalias{labbe2022} lie at or just beyond the stellar mass density constraint in \lcdm.

There are several sources of observational uncertainty that enter these results. The flux calibration of NIRCam is continually being updated; \citetalias{labbe2022} use calibrations that take into account updated detector offsets that are not yet part of the official \jwst\ reduction pipeline (see, e.g., \citealt{Boyer2022} for examples of this effect and \citealt{nardiello2022} for related discussions of empirical point spread function modeling for \jwst). With NIRCam photometry, a Balmer or $4000$~\AA\ break at $z \sim 5$ can be mistaken for a Lyman-$\alpha$ break at $z \ga 12$ \citep{zavala2023}; the \citetalias{labbe2022} sample was selected to contain both Lyman \textit{and} Balmer breaks, however, and is at low enough redshift (relative to $z\sim 15$ sources) that NIRCam filters can typically exclude $z\sim 5$ photometric solutions. The resulting photometric redshift estimates have single, narrow ($\sigma_{z}\approx 0.25$) peaks. The masses of the galaxies are computed using the median of four methods for fitting the photometry (see \citetalias{labbe2022} for details) and assume a \citet{salpeter1955} IMF. Different assumptions about the photometry (in particular, properties of nebular emission lines) or IMF could affect the derived stellar masses, with the latter being a particularly intriguing possibility. The mass of the candidate at $z \sim 7.5$ was also corrected for the possibility of amplification by mild gravitational lensing; this effect is estimated by \citetalias{labbe2022} to be $0.15$~dex, and the reported mass (and stellar mass density) of this object are therefore reduced by this amount to compensate. The error bars in Figure~\ref{fig:rho_star} include errors in the volume estimates coming from both sample variance and Poisson noise, with the latter always being dominant in the regime considered here \citep{trenti2008, behroozi2018}.

The discrepancy between the observed high-redshift galaxy candidates and \lcdm\ expectations is robust to uncertainties in cosmological parameters in the base \lcdm\ model: the precision on each of the relevant parameters is at the $\lesssim 1\%$ level \citep{planck2020}. Intriguingly, extensions to the base \lcdm\ with enhanced values of $\sigma_8$ and the physical matter density $\Omega_{\rm m}h^2$ --- such as some Early Dark Energy (EDE) models whose aim is to resolve the Hubble Tension --- predict earlier structure formation and a higher abundance of halos at fixed mass at high redshift \citep{klypin2021}, which would enhance the baryonic reservoirs available for forming early massive galaxies. Taking the best-fit EDE parameters from \citet{smith2022}, the cumulative comoving baryonic density contained in halos more massive than $\mhalo=\mstar/\fbary$ for the most massive \citetalias{labbe2022} galaxy candidate at $z \approx 9.1$ is a factor of 3.3 larger in EDE than in base \lcdm, which is non-negligible; the \citetalias{labbe2022} data points would then lie at $\epsilon=0.72$ instead of $\epsilon=0.99$. However, this EDE cosmology is in stronger tension with values of $S_8=\sigma_8\,\sqrt{\Omega_{\rm m}/0.3}$ measured at low redshift and predict that the Universe is $\approx 13$ billion years old (as opposed to $13.8$ billion years in the base \lcdm\ model), which is in moderate tension with the measured ages of ultra-faint galaxies and globular clusters \citep{boylan-kolchin2021}. 

At the redshifts studied here, $z \approx 7-10$, the Sheth-Tormen mass function overestimates the abundance of massive halos by $20-50\%$ relative to numerical simulations \citep{reed2003, despali2016, shirasaki2021, wang2022a}, meaning their true abundance at high redshift is likely lower than the Sheth-Tormen prediction and the constraints derived here are conservative. However, the lack of detailed comparisons between theory and simulations at high redshifts and high masses points to the importance of continued theoretical work in understanding the universality and applicability of halo mass function parameterizations in regimes relevant for \jwst\ observations (and other forthcoming observatories).

The tension discussed in this paper is straightforward: the masses measured by \citetalias{labbe2022} are only consistent with expectations from the standard cosmological model at the reported redshifts if star formation in the earliest phases of galaxy formation is incredibly efficient ($\epsilon \ge 0.57$). In the low-redshift Universe, such efficiencies are never seen, with $\epsilon \lesssim 0.2$ for all galaxies. The theoretical expectation is that efficiencies do indeed increase at high redshift \citep{tacchella2018}, though $\epsilon \ga 0.57$ is still highly extreme and likely implausibly high. If the explanation of the \citetalias{labbe2022} galaxies is indeed a very high star formation efficiency, it implies that the star formation histories of such systems must rise steeply with time, following the behavior of the baryon reservoirs inside of virialized structures in \lcdm. The results presented here could also be explained if the stellar IMF differs substantially from the assumed Salpeter form, the flux calibration of NIRCam changes from the latest post-flight determinations, or the volumes currently surveyed turn out to be highly atypical. 

If none of these explanations holds up and these massive galaxies are spectroscopically confirmed, they will pose a serious challenge for \lcdm\ structure formation with parameters given by \citet{planck2020} because they signify the existence of a larger reservoir of collapsed baryons than is possible in this model. Forthcoming wider-field \jwst\ surveys, along with \jwst\ spectroscopy of massive galaxy candidates, should be able to quickly confirm or refute the existence of this tension. Furthermore, the compatibility of any additional high-redshift galaxies or galaxy candidates discovered in \jwst\ observations with \lcdm\ expectations can be assessed in a straightforward way via Fig.~\ref{fig:fig1}. If analysis of \jwst\ data continues to reveal the presence of strikingly massive galaxies at very early cosmic epochs, more exciting surprises lie ahead for the fields of galaxy formation and cosmology.

\vspace{-0.2cm} 
\section*{Data Availability}
Data from \citetalias{labbe2022}, including $\mstar$ estimates and photometric redshifts, are available at \url{https://github.com/ivolabbe/red-massive-candidates}; this paper uses data from {\tt sample\_revision3\_2207.12446.ecsv}, commit {\tt 59fbbfa} (from 2023.01.02). All calculations that go into the figures in this paper will be made publicly available at \url{https://github.com/mrbk/JWST_MstarDensity}.

\vspace{-0.2cm} 
\section*{Acknowledgments} 
This paper is dedicated to the memory of Steven Weinberg, who would have been thrilled to see how well \jwst\ is working and excited to learn what it will reveal about cosmology and galaxy formation across a variety of cosmic epochs. I thank Pieter van Dokkum and Ivo Labb\'e for sharing data from \citetalias{labbe2022}, Steve Finkelstein, Pawan Kumar, and Dan Weisz for helpful discussions, and the referees of the paper for providing helpful comments that improved the clarity of this paper. 
I acknowledge support from the University of Texas at Austin through the Faculty Research Assignment program, NSF CAREER award AST-1752913, NSF grants AST-1910346 and AST-2108962, NASA grant 80NSSC22K0827, and HST-AR-15809, HST-GO-15658, HST-GO-15901, HST-GO-15902, HST-AR-16159, and HST-GO-16226 from the Space Telescope Science Institute, which is operated by AURA, Inc., under NASA contract NAS5-26555. I am very grateful to the developers of the python packages that I used in preparing this paper: {\sc numpy} \citep{numpy2020}, {\sc scipy} \citep{scipy2020}, {\sc matplotlib} \citep{matplotlib}, {\sc hmf} \citep{murray2013, murray2014}, and {\sc ipython} \citep{ipython}. This research has made extensive use of NASA’s Astrophysics Data System (\url{http://adsabs.harvard.edu/}) and the arXiv e-Print service (\url{http://arxiv.org}).

\vspace{-0.2cm} 
\bibliography{draft.bbl}

\end{document}